\documentclass[12pt]{article}
\usepackage{graphicx, amsmath, amssymb, color}
\usepackage[numbers, sort&compress]{natbib}
\usepackage{fullpage}

\newcommand{\F}{\mathcal{F}}
\newcommand{\beq}{\begin{equation}}
\newcommand{\eeq}{\end{equation}}

\newcommand{\s}{\sigma}

\newcommand{\fub}{f_\text{ub}}
\newcommand{\fuf}{f_\text{uf}}
\newcommand{\tbar}{\bar{t}}
\newcommand{\lbar}{\bar{\ell}}

\newcommand{\Spath}{S_\text{path}}
\newcommand{\Scom}{S_\text{com}}

\newcommand{\EOb}{E_b^\text{min}}
\newcommand{\EObone}{E_{b_1}^\text{min}}
\newcommand{\EObtwo}{E_{b_2}^\text{min}}
\newcommand{\EOf}{E_f^\text{ref}}

\begin{document}

\title{Protein folding and binding can emerge as evolutionary spandrels through structural coupling}
\author{Michael Manhart$^1$ and Alexandre V. Morozov$^{1,2}$\footnote{Corresponding author: \texttt{morozov@physics.rutgers.edu}} \\ 
	\small{\emph{$^1$Department of Physics and Astronomy and $^2$BioMaPS Institute for Quantitative Biology,}} \\
	\small{\emph{Rutgers University, Piscataway, NJ 08854, USA}}
	}
\date{}
\maketitle


\begin{abstract}
Binding interactions between proteins and other molecules mediate numerous cellular processes, including metabolism, signaling, and regulation of gene expression.  These interactions evolve in response to changes in the protein's chemical or physical environment (such as the addition of an antibiotic), or when genes duplicate and diverge.
Several recent studies have shown the importance of folding stability in constraining protein evolution. 
Here we investigate how structural coupling between protein folding and binding -- the fact that most proteins can only bind their targets when folded -- gives rise to evolutionary coupling between the traits of folding stability and binding strength.
Using biophysical and evolutionary modeling, we show how these protein traits can emerge as evolutionary ``spandrels'' even if they do not confer an intrinsic fitness advantage. In particular, proteins can evolve strong binding interactions that have no functional role but merely serve to stabilize the protein if misfolding is deleterious.  Furthermore, such proteins may have divergent fates, evolving to bind or not bind their targets depending on random mutation events.  These observations may explain the abundance of apparently nonfunctional interactions among proteins observed in high-throughput assays.
In contrast, for proteins with both functional binding and deleterious misfolding, evolution may be highly predictable at the level of biophysical traits: adaptive paths are tightly constrained to first gain extra folding stability and then partially lose it as the new binding function is developed.
These findings have important consequences for our understanding of fundamental evolutionary principles of both natural and engineered proteins.
\end{abstract}







Proteins carry out a diverse array of chemical and mechanical functions in the cell, ranging from metabolism to signaling~\cite{Creighton1992}.  Thus proteins serve as central targets for natural selection in wild populations, as well as a key toolbox for bioengineering novel molecules with medical and industrial applications~\cite{Campbell2002, Bloom2009}.
Most proteins must fold into their native state, a unique three-dimensional conformation, in order to perform their function, which typically involves binding a target molecule such as DNA, RNA, another protein, or a small ligand~\cite{Creighton1992}. Misfolded proteins may also form toxic aggregates and divert valuable protein synthesis and quality control resources~\cite{Bucciantini2002, Drummond2008, GeilerSamerotte2011, Bershtein2013}.  It is therefore imperative that the folded state be stable against the thermal fluctuations present at physiological temperatures.  However, biophysical experiments and computational studies reveal that most random mutations in proteins destabilize the folded state~\cite{Tokuriki2007, Tokuriki2008}, including mutations that improve function~\cite{Wang2002, Tokuriki2008, Sun2013}.  As a result many natural proteins tend to be only marginally stable, mutationally teetering at the brink of substantial unfolding~\cite{Taverna2002, Zeldovich2007}.  With proteins in such a precarious evolutionary position, how can they evolve new functions while maintaining sufficient folding stability?
     

     Directed evolution experiments have offered a window into the dynamics of this process~\cite{Campbell2002, Bloom2009}, indicating the importance of compensatory mutations, limited epistasis, and mutational robustness.  Theoretical efforts to describe protein evolution in biophysical terms have focused on evolvability~\cite{Bloom2006}, global properties of protein interaction networks~\cite{Johnson2011, Heo2011}, and reproducing observed distributions of protein stabilities and evolutionary rates~\cite{DePristo2005, Drummond2005, Zeldovich2007, Serohijos2012}.
However, a subtle but key property of proteins has not been explored in this context: structural coupling of folding and binding (the fact that folding is required for function) implies evolutionary coupling of folding stability and binding strength.  Thus selection acting directly on only one of these traits may produce apparent, indirect selection for the other.  The importance of this effect was popularized by Gould and Lewontin in their influential paper on evolutionary ``spandrels''~\cite{Gould1979}, defined as traits that evolve as byproducts in the absence of direct selection.
Since then the importance of coupling between traits has been explored in many areas of evolutionary biology~\cite{Pigliucci2000}, including various molecular examples~\cite{Taverna2002, Weiss2002, Barrett2011}.  
     
     How do coupled traits affect protein evolution?  We consider a simple model that describes evolution of a new binding interaction in the context of a directed evolution experiment~\cite{Bloom2009}, as a result of gene duplication and divergence~\cite{Tawfik2010}, or in response to a change in the protein's chemical or physical environment, including availability and concentrations of various ligands~\cite{Weinreich2006, Chou2011} as well as temperature~\cite{Wichman1999, Bershtein2012}.  We postulate a fitness landscape as a function of two biophysical traits: stability and the free energy of binding a target molecule.
We then use an exact numerical algorithm~\cite{Manhart2013, Manhart2014} to quantitatively characterize adaptation on this fitness landscape, addressing key evolutionary questions of epistasis~\cite{Poelwijk2007, Szendro2013}, predictability~\cite{Gould1990, Weinreich2006, Lobkovsky2012}, and the tempo of adaptation~\cite{Kimura1983, DePristo2005}.



\section*{Results}


     \textbf{Model of protein energetics.}  We consider a protein with two-state folding kinetics~\cite{Creighton1992}.  In the folded state, the protein has an interface that binds a target molecule.  
Because the protein can bind \emph{only} when it is folded, the binding and folding processes are structurally coupled.
Under the thermodynamic equilibrium assumption (valid when protein folding and binding are faster than typical cellular processes), the probabilities of the three structural states -- folded and bound ($p_\text{f,b}$), folded and unbound ($p_\text{f,ub}$), and unfolded and unbound ($p_\text{uf,ub}$) -- are given by their Boltzmann weights:

\beq
\begin{array}{|c|c|c|}
\hline
\text{State} & \text{Free energy} & \text{Probability} \\
\hline
& & \\
\text{folded,}     & E_f + E_b    & p_\text{f,b} = {Z^{-1}} e^{-\beta(E_f + E_b)} \\
\text{bound} &&\\
&&\\
\text{folded,}   & E_f          & p_\text{f,ub} = {Z^{-1}} e^{-\beta E_f} \\
\text{unbound} &&\\
&&\\
\text{unfolded,} & 0            & p_\text{uf,ub} = {Z^{-1}} \\
\text{unbound} &&\\
&&\\
\hline
\end{array}
\label{eq:protein_states}
\eeq

\noindent Here $\beta$ is the inverse temperature, $E_f$ is the free energy of folding (also known as $\Delta G$), and $E_b = E'_b - \mu$, where $E'_b$ is the binding free energy and $\mu$ is the chemical potential of the target molecule.  For simplicity, we will refer to $E_b$ as the binding energy.  Note that $E_f < 0$ for intrinsically-stable proteins and $E_b < 0$ for favorable binding interactions.  The partition function is $Z = e^{-\beta(E_f + E_b)} + e^{-\beta E_f} + 1$.

     The folding and binding energies depend on the protein's genotype (amino acid sequence) $\s$.  We assume that adaptation only affects ``hotspot'' residues at the binding interface~\cite{Clackson1995, Moreira2007}; the rest of the protein does not change on relevant time scales because it is assumed to be already optimized for folding. If positions away from the binding interface can accept stabilizing mutations (and are not functionally constrained), they may be explicitly included into the model as ``folding hotspots.''  In the present study we focus on $L$ binding hotspot residues which, to a first approximation, make additive contributions to the total folding and binding free energies~\cite{Wells1990} (see SI Methods for the
discussion of non-additive effects):
     
\beq
E_f(\s) = \EOf + \sum_{i = 1}^L \epsilon_f(i, \s^i), \, E_b(\s) = \EOb + \sum_{i = 1}^L \epsilon_b(i, \s^i),
\label{eq:Ef:Eb}
\eeq

\noindent where $\epsilon_{f}(i, \s^i)$ and $\epsilon_{b}(i, \s^i)$ capture the energetic contributions of amino acid $\s^i$ at position $i$.  The reference energy $\EOf$ is the fixed contribution to the folding energy from all other residues in the protein.  Furthermore, by construction it is also the total folding energy of a reference sequence $\s_\text{ref}$ (see Methods), so that each $\epsilon_{f}(i, \s^i)$ can be interpreted as the change in total folding free energy $E_f$ ($\Delta \Delta G$ value) resulting from a single-point mutation of $\s_\text{ref}$.  The parameter $\EOb$ is the minimum binding energy among all genotypes (see Methods). Amino acid energies $\epsilon_{f}(i, \s^i)$ and $\epsilon_{b}(i, \s^i)$ are randomly sampled from distributions constructed using available $\Delta\Delta G$ data and other biophysical considerations (see Methods); the exact shape of these distributions is unimportant for large enough $L$ due to the central limit theorem.


     \textbf{Fitness landscape.}  We construct a simple fitness landscape based on the molecular traits $E_f$ and $E_b$.  Without loss of generality, we assume that the protein contributes fitness 1 to the organism if it is always folded and bound. Let $\fub, \fuf \in [0, 1]$ be the multiplicative fitness penalties for being unbound and unfolded, respectively: the fitness is $\fub$ if the protein is unbound but folded, and $\fub\fuf$ if the protein is both unbound and unfolded.  Then the fitness of the protein averaged over all three possible structural states in Eq.~\ref{eq:protein_states} is given by

\beq
\F(E_f, E_b) = p_\text{f,b} + \fub p_\text{f,ub} + \fub\fuf p_\text{uf,ub}.
\label{eq:fitness}
\eeq

\begin{figure}[ht!]
\centering\includegraphics[scale=1.0]{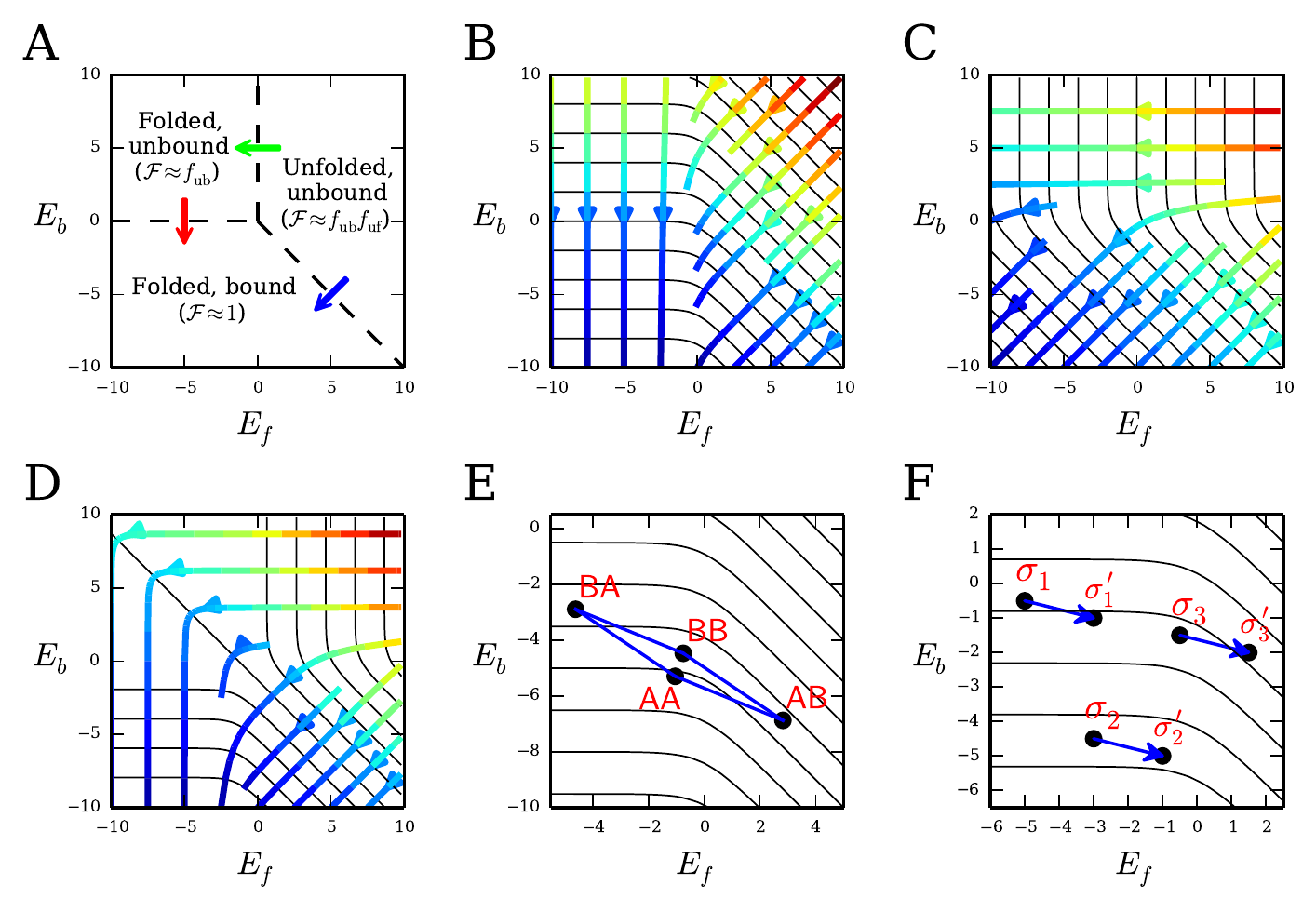}
\caption{
\textbf{Fitness, selection, and epistasis in energy trait space.}
(A)~Phase diagram of protein structural states.  Dashed lines separate structural phases of the protein corresponding to plateaus on the fitness landscape; arrows represent the folding transition (green), binding transition (red), and the coupled folding-binding transition (blue). 
Fitness landscapes $\F(E_f, E_b)$ with direct selection   
(B)~for binding only ($\fub = \fuf = 0$), 
(C)~for folding only ($\fub = 1$, $\fuf = 0$), and 
(D)~for both binding and folding ($\fub = 0.9$, $\fuf = 0$). 
Black contours indicate constant fitness values.  The contours are uniformly spaced in energy space; fitness differences between adjacent contours are not all equal.  Streamlines indicate the direction of the selection ``force'' $\vec{\nabla}\log\F$, with color showing its magnitude (decreasing from red to blue). 
(E)~Projection of a genotype distribution and mutational network into energy space for $L=2$ and a two-letter ($k=2$) alphabet.  
(F)~Blue arrows indicate the same mutation on different genetic backgrounds. When the fitness contours are straight, the mutation is beneficial regardless of the background ($\s_1$ or $\s_2$). However, with curved contours, the same mutation can become deleterious ($\s_3 \to \s_3'$), indicative of sign epistasis.  Sign epistasis from curved contours can give rise to multiple local fitness maxima (e.g., $\mathsf{AA}$ and $\mathsf{BB}$ in (E)).
\label{fig:contours}
}
\end{figure}

\noindent This fitness landscape is divided into three nearly-flat plateaus corresponding to the three protein states of Eq.~\ref{eq:protein_states}, separated by steep thresholds corresponding to the folding and binding transitions (Fig.~\ref{fig:contours}A).  The heights of the plateaus are determined by the values of $\fub$ and $\fuf$, leading to three qualitative regimes of the global landscape structure (Fig.~\ref{fig:contours}B--D). 

     In the first case (Fig.~\ref{fig:contours}B), a protein that is perfectly folded but unbound has no fitness advantage over an unbound and unfolded protein: $\fub = \fub\fuf$.  Thus selection acts directly only on the binding trait.  This regime requires that either $\fub = 0$ (binding is essential, e.g., in the context of conferring antibiotic resistance to the cell~\cite{Weinreich2006}) or $\fuf = 1$ (misfolded proteins are not toxic). The latter case also includes directed evolution experiments where only function is artificially selected for \emph{in vitro}.
In contrast, when $\fub = 1$ and $0 \leq \fuf < 1$ (Fig.~\ref{fig:contours}C), a perfectly folded and bound protein has no fitness advantage over a folded but unbound protein, and thus this case entails direct selection only for folding.  These proteins are harmful to the cell in the misfolded state (e.g., due to aggregation or significant costs of degrading unfolded proteins~\cite{Bucciantini2002, Drummond2008, GeilerSamerotte2011, Bershtein2013}), while binding provides no intrinsic fitness advantage (the protein may have other, functional binding interfaces).
Finally, it is also possible to have distinct selection pressures on both binding and folding.  This occurs when $0 < \fub < 1$ and $0 \leq \fuf < 1$ (Fig.~\ref{fig:contours}D).

     It is straightforward to generalize our three-state model to proteins with additional structural states (other local minima on the folding energy landscape, other binding modes) and allow for simultaneous adaptation at multiple binding interfaces.  Furthermore, the fitness landscape in Eq.~\ref{eq:fitness} can be made an arbitrary nonlinear function of state probabilities.  However, these more complex scenarios would still share the essential features of our basic model: coupling between folding and binding traits and sharp fitness thresholds between bound/unbound and folded/unfolded states.
Thus our qualitative conclusions do not depend on the specific model in Eq.~\ref{eq:fitness}.


     \textbf{Epistasis and local maxima.}  For protein sequences of length $L$ with an alphabet of size $k$, each of the $k^L$ possible genotypes is projected onto the two-dimensional trait space of $E_f$ and $E_b$ (Eq.~\ref{eq:Ef:Eb}) and connected to $L(k-1)$ immediate mutational neighbors, forming a network of states that the population must traverse (a simple example is shown in Fig.~\ref{fig:contours}E).  Adaptive dynamics are determined by the interplay between the structure of the fitness landscape and the distribution of genotypes in trait space.

     This interplay gives rise to the possibility of epistasis and multiple local fitness maxima.  Our model is non-epistatic in energy space (Eq.~\ref{eq:Ef:Eb}). When the fitness contours are straight parallel lines, there can be no sign epistasis on the fitness landscape (Fig.~\ref{fig:contours}F).  Magnitude epistasis, on the other hand, is widespread due to the nonlinear dependence of fitness on folding and binding energies.  Curved fitness contours, which occur near folding or binding thresholds in our model (Fig.~\ref{fig:contours}B--D), can produce sign epistasis in fitness, giving rise to multiple local fitness maxima in the genotype space (Fig.~\ref{fig:contours}E).


     \textbf{Evolutionary dynamics.}  We assume that a population encoding the protein of interest evolves in the monomorphic limit: $LNu\log N \ll 1$, where $L$ is the number of residues, $N$ is an effective population size, and $u$ is the per-residue probability of mutation per generation~\cite{Champagnat2006} (see SI Methods).  In this limit, the entire population has the same genotype at any given time, and the rate of substitution from the current genotype to one of its mutational neighbors is given by Eq.~1 in SI Methods. 
We use the strong-selection limit of the substitution rate (Eq.~2 in SI Methods), 
in which the effective population size enters only as an overall time scale.
In this regime, deleterious mutations never fix and adaptive paths have a finite number of steps, terminating at a global or local fitness maximum.  For compact genomic units such as proteins, the monomorphic condition is generally met in multicellular species, although it may be violated in some unicellular eukaryotes and prokaryotes~\cite{Lynch2007_book}.  Sequential fixation of single mutants is also a typical mode of adaptation in directed evolution experiments~\cite{Bloom2009}.  For simplicity, we neglect more complex mutational moves such as indels and recombination.

     Far from the binding and folding thresholds the fitness landscape becomes flat (Fig.~\ref{fig:contours}A) and the strong-selection assumption may be violated. 
To establish the limits of validity for our model, we calculate average selection coefficients of accessible substitutions (defined as $s = \F_\text{final}/\F_\text{initial} - 1$, where $\F_\text{initial}$ and $\F_\text{final}$ are the initial and final fitness values of a substitution), both throughout the landscape and at the local maxima (Fig.~S1).
We observe that for typical values of the effective population size $N \in (10^4, 10^7)$~\cite{Lynch2007_book, Charlesworth2009}, the selection strengths in the model justify our strong-selection approximation for realistic choices of energy parameters.


     \textbf{Quantitative description of adaptation.}  Although our model is valid for any adaptive process, for concreteness we focus on a specific but widely-applicable scenario.  A population begins as perfectly adapted to binding an original target molecule characterized by an energy matrix $\epsilon_{b_1}$ with minimum binding energy $\EObone$ (defining a fitness landscape $\F_1$).
The population is then subjected to a selection pressure which favors binding a new target, with energy matrix $\epsilon_{b_2}$ and minimum binding energy $\EObtwo$ (fitness landscape $\F_2$).
The adaptive paths are first-passage paths leading from the global maximum on $\F_1$ to a local or global maximum on $\mathcal{F}_2$, with fitness increasing monotonically along each path.

     Each adaptive path $\varphi$ with probability $\Pi[\varphi]$ is a sequence of genotypes connecting initial and final states.
Using an exact numerical algorithm (SI Methods)~\cite{Manhart2013, Manhart2014},
we determine the path-length distribution $\rho(\ell)$, which gives the probability of taking an adaptive path with $\ell$ amino acid substitutions, and the mean adaptation time $\tbar$.  We also introduce $\Spath$, the entropy of the adaptive paths:
     
\beq
\Spath = - \sum_{\varphi} \Pi[\varphi] \log \Pi[\varphi].
\eeq

\noindent 
The path entropy is maximized when evolution is neutral, resulting in all paths of a given length being accessible and equally likely: $\Spath = \lbar \log L(k-1)$~\cite{Manhart2014}, where $\lbar$ is the average path length.

     We also consider the path density $\psi(\s)$, which gives the total probability of reaching a state $\s$ at any point along a path.  When $\s$ is a final state (a local fitness maximum on $\F_2$), the path density is equivalent to the commitment probability.  We calculate the entropy $\Scom$ of the commitment probabilities as

\beq
\Scom = - \sum_{\text{final states } \s} \psi(\s) \log \psi(\s).
\eeq



     \textbf{Direct selection for binding only.}  We first focus on the $\fub = \fuf\fub$ case in Eq.~\ref{eq:fitness}. 
The geometry of the fitness contours is invariant under overall shifts in the binding energy $E_b$ (Fig.~\ref{fig:contours}B); equivalently, the direction (but not the magnitude) of the selection force ($\vec{\nabla}\log\F/|\vec{\nabla}\log\F|$) does not depend on $E_b$. 
Thus without loss of generality, we set $\EObone = \EObtwo$ in this section.  The contours of constant fitness are parallel to the $E_f$ axis when $E_f$ is low, indicating that, as expected, selection acts only on binding when proteins are sufficiently stable.

     However, for marginally stable proteins~\cite{Taverna2002, Kumar2006, Zeldovich2007}, the fitness contours begin to curve downward,
indicating apparent, indirect selection for folding, even though selection acts directly only on the binding trait.
Thus, adaptation will produce a trait (more stability) that is neutral at the level of the fitness function simply because it is coupled with another trait (binding) that is under selection.
Folding stability can therefore be considered an evolutionary spandrel~\cite{Gould1979}. Proteins may even be intrinsically unstable ($E_f > 0$) and only fold when bound ($E_f + E_b < 0$), which we refer to as binding-mediated stability~\cite{Dixit2013}.  In this regime, the fitness contours approach diagonal lines: selection effectively acts to improve both binding and folding equally (Fig.~\ref{fig:contours}B).

     An example realization of evolutionary dynamics in the marginally stable regime is shown in Fig.~\ref{fig:case1}A,B (see Fig.~S2 for stable and intrinsically unstable examples, and Fig.~S3 for averaged distributions of initial, intermediate, and final states).  There is typically just one or two fitness maxima;
all maxima are usually accessible
(Fig.~\ref{fig:case1}C).
For stable proteins, the global maximum almost always coincides with the best-binding genotype and is usually as far as a randomly-chosen genotype from the best-folding genotype (Fig.~\ref{fig:case1}D; two random sequences are separated by $1 - 1/k = 0.8$ for $k=5$).  However, as $E_f$ becomes greater, the average distance between the maxima and the best-binding genotype increases, while the average distance between the maxima and the best-folding genotype decreases, until they meet halfway for intrinsically unstable proteins, where effective selection for binding and folding is equally strong (Fig.~\ref{fig:case1}D).  In general the maxima lie on or near the Pareto front~\cite{Shoval2012}, defined here as the set of genotypes such that either $E_f$ or $E_b$ cannot be decreased further without increasing the other (the global maximum is always on the front, while local maxima may not be) (Fig.~\ref{fig:case1}A, Fig.~S2).

     As $E_f$ increases, the average distance between initial and final states for adaptation decreases.  As a result the average path length (number of substitutions) decreases as well, although the variance of path lengths is relatively constant over all energies (Fig.~\ref{fig:case1}E).  The path entropy per substitution $\Spath/\lbar$ also decreases with $E_f$, reflecting greater constraints on adaptive paths (note that $\Spath/\lbar = \log L(k-1) \approx 3.2$ for neutral evolution). 
Finally, $\Scom \approx 0.31$ in the marginally stable regime (Fig.~\ref{fig:case1}F). Since the average number of maxima is $\approx 1.9$ in this regime (Fig.~\ref{fig:case1}C), the maximum value of $\Scom$ is $\log 1.9 \approx 0.64$, indicating that not all maxima are equally accessible.


     \textbf{Direct selection for folding only.} In this regime, $\fub = 1$ and $0 \leq \fuf < 1$ in Eq.~\ref{eq:fitness}.
Similar to the previous case, the geometry of the fitness contours and thus most landscape properties are now independent of $E_f$ (Fig.~\ref{fig:contours}C); equivalently, normalized selection force $\vec{\nabla}\log\F/|\vec{\nabla}\log\F|$ does not depend on $E_f$.

     When the nonfunctional binding is weak, the fitness contours are parallel to the $E_b$ axis, indicating that selection acts only on folding (Fig.~\ref{fig:contours}C).  However, with increasing binding strength the fitness contours curve such that the effective selection force attempts to improve both binding and folding equally.  Thus binding emerges as an evolutionary spandrel
in this case.
The weak-binding regime yields a single fitness maximum due to the lack of sign epistasis; this maximum predominantly coincides with the best-folding genotype (Fig.~\ref{fig:case2}A).  However, once the binding interaction becomes stronger, there is an increased likelihood of multiple local maxima, located between the best-folding and best-binding genotypes.  

     


     Depending on the abundance of the old and new ligands in the cell and their binding properties, several adaptive scenarios may take place.  First, the best-binding strengths $\EObone$ and $\EObtwo$ of the old and new targets may be similar in magnitude.  If both are weak, initial and final states are likely to be the best-folding genotype or close to it (Fig.~\ref{fig:case2}A); in this case, there is a high probability that no adaptation will occur (Fig.~\ref{fig:case2}B).  When $\EObone$ and $\EObtwo$ are both low, adaptation usually occurs to accommodate the binding specificity of the new ligand (Fig.~\ref{fig:case2}B, Fig.~S4A).
Surprisingly, we see that proteins frequently evolve stronger binding at the expense of folding (bottom panel of Fig.~S4A).  This happens due to the constraints of the genotype-phenotype map: not enough genotypes are available to optimize both traits simultaneously.

     It is also possible to gain or lose binding affinity at the nonfunctional interface through adaptation.  In the first case, the new target has stronger binding than the old one ($\EObtwo < \EObone$).  Thus the initial state is the best-folding genotype or close to it, and the protein adapts toward a genotype with intermediate folding and binding (Fig.~S4B).  As before, adaptation is tightly constrained by the genotype-phenotype map, sacrificing the trait (folding stability) under direct selection in order to affect the spandrel (nonfunctional binding interaction).  Effectively, the protein switches from being ``self-reliant'' to needing a binding partner.  In the second case ($\EObone < \EObtwo$), the dynamics is opposite:
the protein loses its nonfunctional binding interface and becomes self-reliant (Fig.~S4C).
Thus proteins may acquire or lose binding interfaces depending on the availability of ligands that can participate in binding-mediated stability.  If the protein's stability becomes suboptimal due to an environmental change, its stability may be restored not only through stabilizing mutations, but also by developing a novel binding interface.


     \textbf{Divergent evolutionary fates.}  In the region where the fitness contours in Fig.~\ref{fig:contours}C are curved, it is possible to have two or more local maxima accessible to adaptation, with at least one having negative $E_b$ (strong binding) and at least one having positive $E_b$ (negligible binding) (see Fig.~\ref{fig:case2}C,D for an example landscape).  The selection streamlines are divergent in this regime (Fig.~\ref{fig:contours}C).
Thus a protein has two fates available to it: one in which it evolves to bind the target and another in which it does not.  The eventual fate of the protein is determined by random mutation events.  Indeed, the distribution of final states is strongly bimodal (Fig.~\ref{fig:case2}E), yielding a sizable probability of divergent fates across a range of binding energies (Fig.~\ref{fig:case2}F).


     \textbf{Simultaneous selection for binding and folding.}  Finally we consider a general case in which $0 < \fub < 1$ and $0 \leq \fuf < 1$ in Eq.~\ref{eq:fitness} (Fig.~\ref{fig:contours}D).
The fitness landscape is divided into two regions by a straight diagonal contour with fitness $\fub$ and slope $-1$. Below this contour, the landscape is qualitatively similar to the case of selection for binding only (Fig.~\ref{fig:contours}B), while above the contour the landscape resembles that of the folding-only selection scenario (Fig.~\ref{fig:contours}C). Thus evolutionary dynamics for proteins with favorable binding and folding energies will largely resemble the case of selection for binding only.  However, a qualitatively different behavior will be observed if the distribution of genotypes straddles the diagonal contour (Fig.~\ref{fig:case3}).  This will occur when initial folding stability is marginal and initial binding is unfavorable.  In this case, selection streamlines around the diagonal contour (Fig.~\ref{fig:contours}D) and the genotype-phenotype map tightly constrain the adaptive paths to gain extra folding stability first, and then lose it as the binding function is improved.


     \textbf{Tempo and rhythm of adaptation.}
The strength of selection is the primary determinant of the average adaptation time $\tbar$.  If the selection coefficient $s$ is small (but $Ns > 1$), the substitution rate $W(\s'|\s)$ in SI Methods Eq.~1 
is proportional to $s$.  Thus, as selection becomes exponentially weaker for lower energies (Fig.~S1), adaptation becomes exponentially slower.
The distribution of the total adaptation time over an adaptive path is highly nonuniform. For example, in the case of selection for binding only and a marginally stable protein, the adaptation time is concentrated at the end of the path, one mutation away from the final state (Fig.~S5A,B).  Substitutions at the beginning of the path occur quickly because there are many possible beneficial substitutions and because selection is strong; in contrast, at the end of the path adaptation slows down dramatically as beneficial mutations are depleted and selection strength weakens.  This behavior is observed in most of the other model regimes as well.

     The exception to this pattern occurs in the case of selection for both binding and folding in marginally-stable and marginally-bound proteins, due to the unique contour geometry (Fig.~\ref{fig:contours}D).  As the adaptive paths wrap around the diagonal contour in the region of high $E_b$ and low $E_f$, the landscape flattens, making selection weaker and substitutions slower (Fig.~S5C).  Thus most of the waiting occurs in the middle of the path rather than the end (Fig.~S5D).
Adaptation accelerates toward the end of the path as the strength of selection increases again.  If the intermediate slow-down is significant enough, a protein may not have time to complete the second half of its path before environmental conditions change, so that it will never evolve the new binding function.


\section*{Discussion}

     \textbf{Protein folding and binding as evolutionary spandrels.}  In the decades since Gould and Lewontin's paper~\cite{Gould1979}, the existence of evolutionary spandrels has emerged as a critical evolutionary concept.
There are many possible scenarios in which spandrels can evolve~\cite{Gould1979, Pigliucci2000}, although two key mechanisms are neutral processes, such as genetic drift and biases in mutation and recombination~\cite{Lynch2007_NatRevGenet}, and indirect selection arising from coupled traits.  Here we have focused on the latter, which we expect to be more important on short time scales.

     It has been previously argued that the marginal stability of most proteins may be an evolutionary spandrel that evolved
due to mutation-selection balance~\cite{Taverna2002, Zeldovich2007, Bloom2009}.  We suggest more broadly that having folding stability at all may be a spandrel for proteins with no misfolding toxicity.  Even more striking is the possibility that some binding interactions may be spandrels that evolved solely to stabilize proteins with toxic misfolding; this would significantly affect our interpretation of data on proteome-wide interactions~\cite{Stark2006}.  In particular, we expect more widespread nonfunctional interactions among proteins with less intrinsic stability. 
Indeed, protein abundance is believed to correlate positively with stability ($-E_f$) to explain the observed negative correlation of abundance with evolutionary rate~\cite{Drummond2005, Serohijos2012}.  Furthermore, models of protein-protein interaction networks imply that protein abundance also correlates negatively with the number of interactions~\cite{Heo2011}.  Together these argue that stability should indeed be negatively correlated with the number of interactions.  Experiments on specific proteins also support this finding: for example, destabilizing mutations in \emph{E. coli} dihydrofolate reductase were found to be compensated at high temperature by protein binding, which protected against toxic aggregation~\cite{Bershtein2012}.  Previously the role of binding-mediated stability has been primarily discussed in the context of intrinsically disordered proteins~\cite{Wright2009}, described by the high $E_f$ regime of our model.

     \textbf{Pareto optimization of proteins.}  The Pareto front is a useful concept in problems of multi-objective optimization~\cite{Shoval2012}.
The Pareto front in our model consists of the protein sequences along the low $E_f$, low $E_b$ edge of the genotype distribution (see e.g. Fig.~\ref{fig:case1}A).
Pareto optimization assumes that all states on the front are valid final states for adaptation; this in turn implies that fitness has linear dependence on the individual traits.  However, nonlinear fitness functions with saturation effects will confound this assumption.
Our model shows how this nonlinearity leads to a small subset of true final states on or even off the front. Thus Pareto optimization does not capture a key feature of the underlying biophysics, providing only a rough approximation to the true dynamics.

     \textbf{Epistasis and evolutionary predictability.}  Our results also shed light on the role of epistasis -- the correlated effects of mutations at different sites -- in protein evolution.  Epistasis underlies the ruggedness of fitness landscapes~\cite{Poelwijk2007, Szendro2013}.
Magnitude epistasis is widespread in our model, while sign epistasis only arises in regions where the fitness contours are curved (Fig.~\ref{fig:contours}E,F). This picture is qualitatively consistent with studies of empirical fitness landscapes~\cite{Szendro2013} and with directed evolution experiments~\cite{Bloom2009}.

     Epistasis determines the predictability of evolution, an issue of paramount importance in biology~\cite{Gould1990, Weinreich2006, Lobkovsky2012}.
In most cases considered here,
limited sign epistasis gives rise to less predictable intermediate pathways (high $\Spath$) but highly predictable final outcomes (low $\Scom$).

     However, there are two major exceptions to this pattern.  First, proteins with a binding interaction under no direct selection may have multiple local maxima, some with strong and others with weak binding (Fig.~\ref{fig:case2}).  Here both intermediate pathways and final states are unpredictable -- pure chance, in the form of random mutations, drives the population to one binding fate or the other.
The second exception occurs in proteins with direct selection for both binding and folding.  Here there is usually a single maximum, but the adaptive paths are tightly constrained in energy space (Fig.~\ref{fig:case3}).
Thus evolution of proteins with both functional binding and deleterious misfolding,
which should include a large fraction of natural proteins,
is highly predictable at the level of energy traits.


	


\section*{Methods}

     \textbf{Energetics of protein folding and binding.}  Folding energetics are probed experimentally and computationally by measuring the changes in $E_f$ resulting from single point mutations.  Since these changes are observed to be universally distributed over many proteins~\cite{Tokuriki2007}, we sample entries of $\epsilon_f$ from a Gaussian distribution with mean 1.25 kcal/mol and standard deviation 1.6 kcal/mol. For the reference sequence $\s_\text{ref}$, $\epsilon_f(i, \s^i_\text{ref}) = 0$ for all $i \in \{1, \ldots, L\}$, such that $E_f(\s_\text{ref}) = \EOf$.  The parameter $\EOb$ is defined as the binding energy of the genotype $\s_\text{bb}$ with the lowest $E_b$: $\epsilon_b(i, \s^i_\text{bb}) = 0$ for all $i \in \{1, \ldots, L\}$.  Since binding hotspot residues typically have a 1--3 kcal/mol penalty for mutations away from the wild-type amino acid
~\cite{Clackson1995, Moreira2007}, we sample the other entries of $\epsilon_b$ from an exponential distribution defined in the range of $(1,\infty)$ kcal/mol, with mean 2 kcal/mol.  This distribution is consistent with alanine-scanning experiments which probe energetics of amino acids at the binding interface~\cite{Thorn2001}.
We consider $L=6$ hotspot residues and a reduced alphabet of $k=5$ amino acids (grouped into negative, positive, polar, hydrophobic, and other), resulting in $5^6 = 15625$ possible genotypes. Our population genetics model and the algorithm for exact calculation of adaptive path statistics are available in SI Methods.

\section*{Acknowledgments}

     A.V.M. acknowledges support from National Institutes of Health (R01 HG004708) and an Alfred P. Sloan Research Fellowship.


\bibliographystyle{pnas2011}
\bibliography{Bibliography}

\begin{thebibliography}{10}

\bibitem{Creighton1992}
Creighton TE (1992) {\em Proteins: Structures and Molecular Properties}.
\newblock (W.H. Freeman and Company, New York).

\bibitem{Campbell2002}
Campbell RE et~al. (2002) A monomeric red fluorescent protein.
\newblock {\em Proc Natl Acad Sci USA} 99:7877--7882.

\bibitem{Bloom2009}
Bloom JD, Arnold FH (2009) In the light of directed evolution: Pathways of
  adaptive protein evolution.
\newblock {\em Proc Natl Acad Sci USA} 106:9995--10000.

\bibitem{Bucciantini2002}
Bucciantini M et~al. (2002) Inherent toxicity of aggregates implies a common
  mechanism for protein misfolding diseases.
\newblock {\em Nature} 416:507--511.

\bibitem{Drummond2008}
Drummond DA, Wilke CO (2008) Mistranslation-induced protein misfolding as a
  dominant constraint on coding-sequence evolution.
\newblock {\em Cell} 134:341--352.

\bibitem{GeilerSamerotte2011}
Geiler-Samerotte KA et~al. (2011) Misfolded proteins impose a dosage-dependent
  fitness cost and trigger a cytosolic unfolded protein response in yeast.
\newblock {\em Proc Natl Acad Sci USA} 108:680--685.

\bibitem{Bershtein2013}
Bershtein S, Mu W, Serohijos AWR, Zhou J, Shakhnovich EI (2013) Protein quality
  control acts on folding intermediates to shape the effects of mutations on
  organismal fitness.
\newblock {\em Mol Cell} 49:133--144.

\bibitem{Tokuriki2007}
Tokuriki N, Stricher F, Schymkowitz J, Serrano L, Tawfik DS (2007) The
  stability effects of protein mutations appear to be universally distributed.
\newblock {\em J Mol Biol} 369:1318--1332.

\bibitem{Tokuriki2008}
Tokuriki N, Stricher F, Serrano L, Tawfik DS (2008) How protein stability and
  new functions trade off.
\newblock {\em PLoS Comput Biol} 4:e1000002.

\bibitem{Wang2002}
Wang X, Minasov G, Shoichet BK (2002) Evolution of an antibiotic resistance
  enzyme constrained by stability and activity trade-offs.
\newblock {\em J Mol Biol} 320:85--95.

\bibitem{Sun2013}
Sun SB et~al. (2013) Mutational analysis of 48g7 reveals that somatic
  hypermutation affects both antibody stability and binding affinity.
\newblock {\em J Am Chem Soc} 135:9980--9983.

\bibitem{Taverna2002}
Taverna DM, Goldstein RA (2002) Why are proteins marginally stable?
\newblock {\em Proteins} 46:105--109.

\bibitem{Zeldovich2007}
Zeldovich KB, Chen P, Shakhnovich EI (2007) Protein stability imposes limits on
  organism complexity and speed of molecular evolution.
\newblock {\em Proc Natl Acad Sci USA} 104:16152--16157.

\bibitem{Bloom2006}
Bloom JD, Labthavikul ST, Otey CR, Arnold FH (2006) Protein stability promotes
  evolvability.
\newblock {\em Proc Natl Acad Sci USA} 103:5869--5874.

\bibitem{Johnson2011}
Johnson ME, Hummer G (2011) Nonspecific binding limits the number of proteins
  in a cell and shapes their interaction networks.
\newblock {\em Proc Natl Acad Sci USA} 108:603--608.

\bibitem{Heo2011}
Heo M, Maslov S, Shakhnovich EI (2011) Topology of protein interaction network
  shapes protein abundances and strengths of their function and nonspecific
  interactions.
\newblock {\em Proc Natl Acad Sci USA} 108:4258--4263.

\bibitem{DePristo2005}
DePristo MA, Weinreich DM, Hartl DL (2005) Missense meanderings in sequence
  space: a biophysical view of protein evolution.
\newblock {\em Nat Rev Genet} 6:678--687.

\bibitem{Drummond2005}
Drummond DA, Bloom JD, Adami C, Wilke CO, Arnold FH (2005) Why highly expressed
  proteins evolve slowly.
\newblock {\em Proc Natl Acad Sci USA} 102:14338--14343.

\bibitem{Serohijos2012}
Serohijos AWR, Rimas Z, Shakhnovich EI (2012) Protein biophysics explains why
  highly abundant proteins evolve slowly.
\newblock {\em Cell Rep} 2:249--256.

\bibitem{Gould1979}
Gould SJ, Lewontin RC (1979) The spandrels of {San Marco} and the {Panglossian}
  paradigm: A critique of the adaptationist programme.
\newblock {\em Proc R Soc Lond B} 205:581--598.

\bibitem{Pigliucci2000}
Pigliucci M, Kaplan J (2000) The fall and rise of {Dr Pangloss}: adaptationism
  and the {Spandrels} paper 20 years later.
\newblock {\em Trends Ecol Evol} 15:66--77.

\bibitem{Weiss2002}
Weiss MA et~al. (2002) Protein structure and the spandrels of {San Marco}:
  Insulin's receptor-binding surface is buttressed by an invariant leucine
  essential for its stability.
\newblock {\em Biochemistry} 41:809--819.

\bibitem{Barrett2011}
Barrett RDH, Hoekstra HE (2011) Molecular spandrels: tests of adaptation at the
  genetic level.
\newblock {\em Nat Rev Genet} 12:767--780.

\bibitem{Tawfik2010}
Soskine M, Tawfik DS (2010) Mutational effects and the evolution of new protein
  functions.
\newblock {\em Nat Rev Genet} 11:572--582.

\bibitem{Weinreich2006}
Weinreich DM, Delaney NF, DePristo MA, Hartl DL (2006) Darwinian evolution can
  follow only very few mutational paths to fitter proteins.
\newblock {\em Science} 312:111--114.

\bibitem{Chou2011}
Chou HH, Chiu HC, Delaney NF, Segr{\`{e}} D, Marx CJ (2011) Diminishing returns
  epistasis among beneficial mutations decelerates adaptation.
\newblock {\em Science} 332:1190--1192.

\bibitem{Wichman1999}
Wichman HA, Badgett MR, Scott LA, Boulianne CM, Bull JJ (1999) Different
  trajectories of parallel evolution during viral adaptation.
\newblock {\em Science} 285:422--424.

\bibitem{Bershtein2012}
Bershtein S, Mu W, Shakhnovich EI (2012) Soluble oligomerization provides a
  beneficial fitness effect on destabilizing mutations.
\newblock {\em Proc Natl Acad Sci USA} 109:4857--4862.

\bibitem{Manhart2013}
Manhart M, Morozov AV (2013) Path-based approach to random walks on networks
  characterizes how proteins evolve new functions.
\newblock {\em Phys Rev Lett} 111:088102.

\bibitem{Manhart2014}
Manhart M, Morozov AV (2014) in {\em First-Passage Phenomena and Their
  Applications}, eds.{} Metzler R, Oshanin G, Redner S.
\newblock (World Scientific, Singapore).

\bibitem{Poelwijk2007}
Poelwijk FJ, Kiviet DJ, Weinreich DM, Tans SJ (2007) Empirical fitness
  landscapes reveal accessible evolutionary paths.
\newblock {\em Nature} 445:383--386.

\bibitem{Szendro2013}
Szendro IG, Schenk MF, Franke J, Krug J, de~Visser JA (2013) Quantitative
  analyses of empirical fitness landscapes.
\newblock {\em J. Stat. Mech.} p. P01005.

\bibitem{Gould1990}
Gould SJ (1990) {\em Wonderful Life: The Burgess Shale and the Nature of
  History}.
\newblock (W. W. Norton and Company, New York, USA).

\bibitem{Lobkovsky2012}
Lobkovsky AE, Koonin EV (2012) Replaying the tape of life: quantification of
  the predictability of evolution.
\newblock {\em Front Gene} 3:246.

\bibitem{Kimura1983}
Kimura M (1983) {\em The Neutral Theory of Molecular Evolution}.
\newblock (Cambridge University Press, Cambridge, UK).

\bibitem{Clackson1995}
Clackson T, Wells JA (1995) A hot spot of binding energy in a hormone-receptor
  interface.
\newblock {\em Science} 267:383--386.

\bibitem{Moreira2007}
Moreira IS, Fernandes PA, Ramos MJ (2007) Hot spots --- a review of the
  protein-protein interface determinant amino-acid residues.
\newblock {\em Proteins} 68:803--812.

\bibitem{Wells1990}
Wells JA (1990) Additivity of mutational effects in proteins.
\newblock {\em Biochemistry} 29:8509--8517.

\bibitem{Champagnat2006}
Champagnat N (2006) A microscopic interpretation for adaptive dynamics trait
  substitution sequence models.
\newblock {\em Stoch Proc Appl} 116:1127--1160.

\bibitem{Lynch2007_book}
Lynch M (2007) {\em The Origins of Genome Architecture}.
\newblock (Sinauer, Sunderland).

\bibitem{Charlesworth2009}
Charlesworth B (2009) Effective population size and patterns of molecular
  evolution and variation.
\newblock {\em Nat Rev Genet} 10:195--205.

\bibitem{Kumar2006}
Kumar MD et~al. (2006) {ProTherm} and {ProNIT}: thermodynamic databases for
  proteins and protein-nucleic acid interactions.
\newblock {\em Nuleic Acids Res} 34:D204--D206.

\bibitem{Dixit2013}
Dixit PD, Maslov S (2013) Evolutionary capacitance and control of protein
  stability in protein-protein interaction networks.
\newblock {\em PLoS Comput Biol} 9:e1003023.

\bibitem{Shoval2012}
Shoval O et~al. (2012) Evolutionary trade-offs, {P}areto optimality, and the
  geometry of phenotype space.
\newblock {\em Science} 336:1157--1160.

\bibitem{Lynch2007_NatRevGenet}
Lynch M (2007) The evolution of genetic networks by non-adaptive processes.
\newblock {\em Nat Rev Genet} 8:803--813.

\bibitem{Stark2006}
Stark C et~al. (2006) {BioGRID}: a general repository for interaction datasets.
\newblock {\em Nucleic Acids Res} 34(Database issue):D535--D539.

\bibitem{Wright2009}
Wright PE, Dyson HJ (2009) Linking folding and binding.
\newblock {\em Curr Opin Struct Biol} 19:31--38.

\bibitem{Thorn2001}
Thorn KS, Bogan AA (2001) {ASE}db: a database of alanine mutations and their
  effects on the free energy of binding in protein interactions.
\newblock {\em Bioinformatics} 17:284--285.

\bibitem{Kimura1962}
Kimura M (1962) On the probability of fixation of mutant genes in a population.
\newblock {\em Genetics} 47:713--719.

\bibitem{Istomin2008}
Istomin AY, Gromiha MM, Vorov OK, Jacobs DJ, Livesay DR (2008) New insight into
  long-range nonadditivity within protein double-mutant cycles.
\newblock {\em Proteins} 70:915--924.

\end{thebibliography}


\newpage
\section*{Figures}

\newpage

\begin{figure}[ht!]
\centering\includegraphics[scale=1.0]{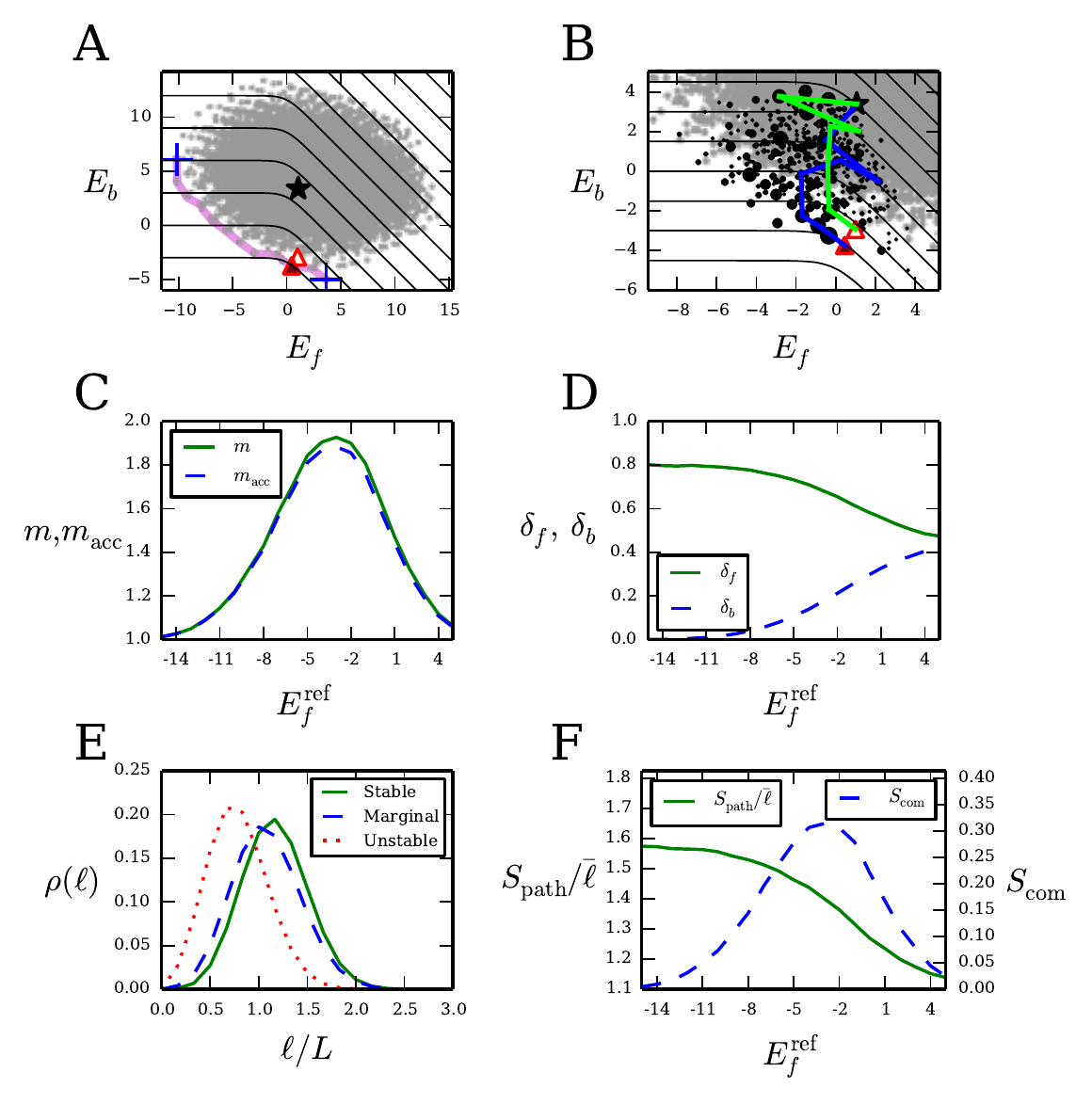}
\end{figure}

\newpage

\begin{figure}[ht!]
\caption{
\textbf{Properties of adaptation with direct selection for binding only.}  
(A)~Global distribution of folding and binding energies for all $k^L = 5^6$ genotypes in a single realization of the model with a marginally stable protein ($\EOf = -3$ kcal/mol).  The black star indicates the initial state for adaptation (global maximum on $\F_1$), red triangles indicate local fitness maxima on $\F_2$, shaded according to their commitment probabilities $\psi(\s)$, and the blue crosses indicate best-folding and best-binding genotypes.  The magenta line connects genotypes on the Pareto front, and the black contours indicate constant fitness $\F_2$. 
(B)~The region of energy space accessible to adaptive paths, zoomed in from (A).  Example paths are shown in blue and green; black circles indicate intermediate states along paths, sized proportional to their path density $\psi(\s)$; small gray circles are genotypes inaccessible to adaptation. 
(C)~Average number $m$ of local fitness maxima (solid, green) and average number $m_\text{acc}$ of local maxima accessible to adaptation (dashed, blue) versus $\EOf$.  The average number of maxima is greatest at $\EOf \approx -3 \text{ kcal/mol}$, where multiple local maxima are separated by $\approx 2.23$ substitutions on average.
(D)~Average per-residue Hamming distance between the maxima and the best-folding genotype ($\delta_f$; solid, green) and the best-binding genotype ($\delta_b$; dashed, blue) versus $\EOf$.
(E)~Average distributions $\rho(\ell)$ of path lengths (number of substitutions) $\ell$ for stable proteins ($\EOf = -15$ kcal/mol), marginally stable proteins ($\EOf = -3$ kcal/mol), and intrinsically unstable proteins ($\EOf = 5$ kcal/mol). 
(F)~Per-substitution path entropy $\Spath/\lbar$ (solid, green) and entropy of commitment probabilities $\Scom$ (dashed, blue)
versus $\EOf$. 
Panel (E) is averaged over $10^5$ realizations of the model; all other averages are taken over $10^4$ realizations. In all panels $\fub = \fuf = 0$ and $\EObone = \EObtwo = -5$ kcal/mol.
\label{fig:case1}
}
\end{figure}

\newpage

\begin{figure}[ht!]
\centering\includegraphics[scale=1.0]{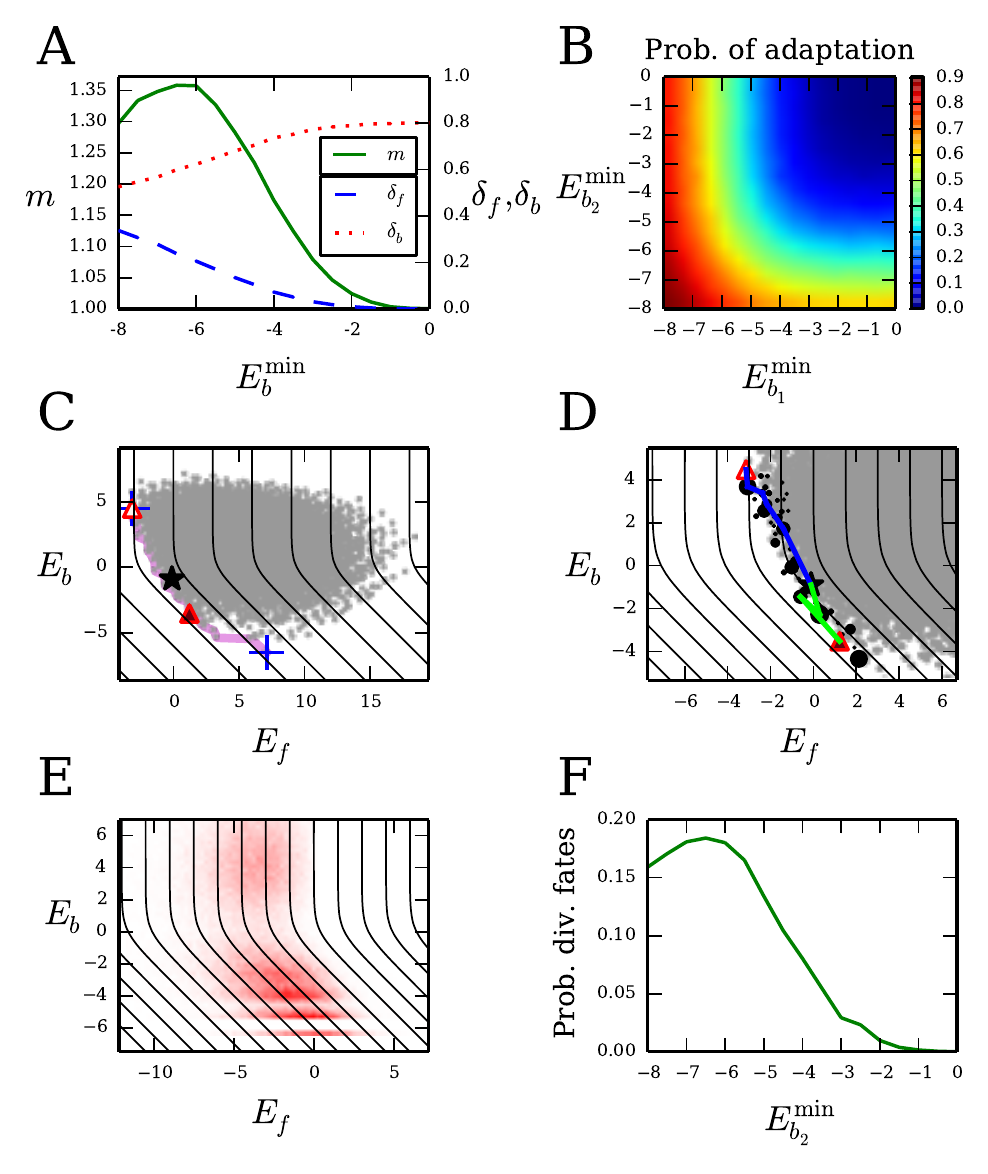}
\caption{
\textbf{Properties of adaptation with direct selection for folding only.} 
(A)~The average number of local maxima $m$ (solid, green) and their average per-residue Hamming distance from the best-folding ($\delta_f$; dashed, blue) and the best-binding ($\delta_b$; dotted, red) genotypes versus $\EOb$.  
(B)~Probability that adaptation occurs when the binding target is changed (i.e., the initial state is not coincident with any of the final states), as a function of $\EObone$ and $\EObtwo$.  
(C,D)~Example landscape with divergent binding fates: there are two accessible local maxima, one with $E_b < 0$ (favorable binding, $\psi(\s) = 0.6$)
and the other with $E_b > 0$ (negligible binding, $\psi(\s) = 0.4$).
All symbols are the same as in Fig.~\ref{fig:case1}A,B.
(E)~Average distribution of local maxima, weighted by their commitment probabilities.  The average commitment entropy for realizations with divergent fates is $\Scom \approx 0.43$.  In (C)--(E) we used $\EObone = \EObtwo = -6.5$ kcal/mol.  
(F)~The probability of having divergent fates versus $\EObtwo = \EObone$.
Panel (E) is averaged over $10^5$ realizations of the model; all other averages are taken over $10^4$ realizations. In all panels $\fub = 1$, $\fuf = 0$, and $\EOf = 0$ kcal/mol.
\label{fig:case2}
}
\end{figure}

\newpage

\begin{figure}[ht!]
\centering\includegraphics[scale=1.0]{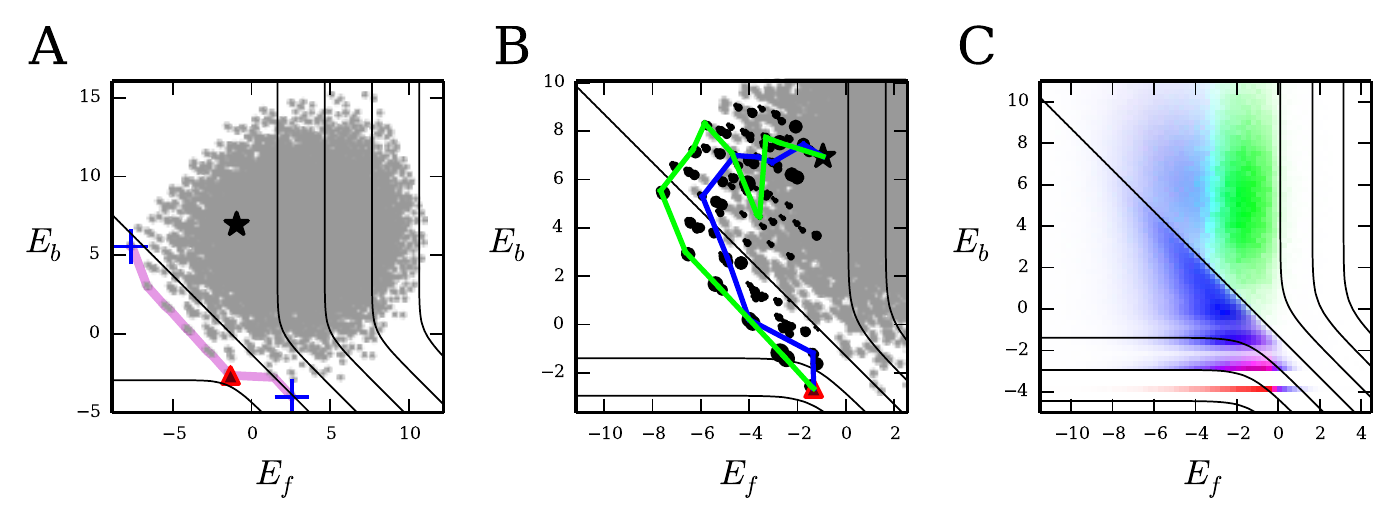}
\caption{
\textbf{Properties of adaptation with direct selection for both folding and binding.}
(A, B)~Distribution of folding and binding energies in an example landscape for a marginally stable and marginally bound protein; all symbols are the same as in Fig.~\ref{fig:case1}A,B. 
(C)~Landscape averaged over $10^5$ realizations.  Distribution of initial states is shown in green, intermediate states in blue (weighted by their path densities), and final states in red (weighted by their commitment probabilities).
In all panels $\fub = 0.9$, $\fuf = 0$, and $\EOf = \EObone = \EObtwo = -4$ kcal/mol.
\label{fig:case3}
}
\end{figure}

\newpage


\begin{center}
\LARGE{Supplementary Material: \\ Protein folding and binding can emerge as evolutionary spandrels through structural coupling}

\vspace{0.5cm}

\large{Michael Manhart$^1$ and Alexandre V. Morozov$^{1,2}$}

\vspace{0.5cm}

	\small{\emph{$^1$Department of Physics and Astronomy and $^2$BioMaPS Institute for Quantitative Biology,}} \\

	\small{\emph{Rutgers University, Piscataway, New Jersey 08854, USA}}
\end{center}


\renewcommand{\theequation}{S\arabic{equation}}
\setcounter{equation}{0}

\section*{Supplementary Methods}

     \textbf{Population genetics model.}  In the monomorphic limit, the population is described by a single point in genotype space~\cite{Champagnat2006}.  The population evolves over time via mutations that arise sequentially and either fix or disappear. Each fixation event leads to an amino acid substitution in the entire population. The rate of making a substitution from genotype $\s$ to genotype $\s'$ is given by~\cite{Kimura1983}

\beq
W(\s'|\s) = Nu~\phi(\s'|\s),
\label{eq:substitution_rate}
\eeq

\noindent where $N$ is the effective population size, $u$ is the mutation rate, and $\phi(\s'|\s)$ is the probability of a single $\s'$ mutant fixing in a population of wild-type $\s$.  Typically the fixation probability depends only on the relative selection coefficient $s = \F(\s')/\F(\s) - 1$ between the two genotypes, where $\F(\s)$ is the fitness of genotype $\s$. 
For example, in the Wright-Fisher model, $\phi(s) = (1 - e^{-2s})/(1 - e^{-2Ns})$, where $N$ is the effective population size~\cite{Kimura1962}.  In the strong-selection limit ($N|s| \gg 1$),

\beq
\phi(s) \approx \left\{
\begin{array}{ll}
1 - e^{-2s} & \text{for $s > 0$} \\
0 & \text{for $s < 0$}
\end{array}
\right.
\label{eq:phi_infinite_N}
\eeq

\noindent Thus the effective population size $N$ sets the overall time scale $(Nu)^{-1}$ of substitutions but does not affect fixation probabilities.


     \textbf{Statistics of adaptive paths.}  We calculate statistical properties of the adaptive paths using a transfer matrix-like algorithm~\cite{Manhart2013, Manhart2014}.  Let $\mathcal{S}$ be the set of all genotypes accessible to adaptation, and let $\mathcal{S}_f$ be the set of final state genotypes (e.g., local fitness maxima).  Define $W(\s'|\s)$ as the rate of making a substitution from genotype $\s$ to genotype $\s'$ (e.g., given by Eq.~S1).  The rate matrix defines $\theta(\s) = (\sum_{\text{nn $\s'$ of $\s$}} W(\s'|\s))^{-1}$, the mean waiting time in genotype $\s$ before a substitution occurs, where the sum is over all genotypes $\s'$ one mutation away from $\s$ (nearest mutational neighbors, ``nn'').  The substitution rates also determine the probability $Q(\s'|\s) = W(\s'|\s) \theta(\s)$ of making the substitution $\s \rightarrow \s'$, given that a substitution occurs out of $\s$.
     
     For each substitution $\ell$ and intermediate genotype $\s$, we calculate $P_\ell(\s)$, the total probability of all paths that end at $\s$ in $\ell$ substitutions; $T_\ell(\s)$, the total average time of all such paths; and $\Gamma_\ell(\s)$, their total entropy.  These quantities obey the following recursion relations:
     
\begin{eqnarray}
P_{\ell}(\s') &=& \sum_{\text{nn } \s \text{ of } \s'} Q(\s'|\s) P_{\ell - 1} (\s), \\
T_{\ell}(\s') &=& \sum_{\text{nn } \s \text{ of } \s'} Q(\s'|\s) \left[ T_{\ell - 1} (\s) + \theta(\s) P_{\ell - 1} (\s) \right], \nonumber \\
\Gamma_{\ell}(\s') &=& \sum_{\text{nn } \s \text{ of } \s'} Q(\s'|\s) \left[ \Gamma_{\ell - 1} (\s) - (\log Q(\s'|\s)) P_{\ell - 1} (\s) \right], \nonumber
\end{eqnarray}

\noindent where $P_0(\s) = 1$ if $\s$ is the initial state and $P_0(\s) = 0$ otherwise, and $T_0(\s) = \Gamma_0(\s) = 0$ for all $\s \in \mathcal{S}$.  The final states $\s \in \mathcal{S}_f$ are treated as absorbing to ensure that only first-passage paths are counted.  We use these transfer matrix objects to calculate the path ensemble quantities described in the text:

\begin{align}
\rho(\ell) & = \sum_{\s \in \mathcal{S}_f} P_\ell(\s), & \psi(\s) & = \sum_{\ell = 1}^\Lambda P_\ell(\s), \\ \nonumber
\tbar = \sum_{\ell=1}^\Lambda \sum_{\s \in \mathcal{S}_f} T_\ell(\s) & = \sum_{\ell=1}^\Lambda \tau(\ell) = \sum_{\s \in \mathcal{S}} \tau(\s), & \tau(\s) & = \sum_{\ell = 1}^\Lambda \theta(\s) P_\ell(\s), \\
\Spath & = \sum_{\ell=1}^\Lambda \sum_{\s \in \mathcal{S}_f} \Gamma_\ell(\s), & \tau(\ell) & = \sum_{\s \in \mathcal{S}} \theta(\s) P_\ell(\s).  \nonumber
\end{align}

\noindent The sums are calculated up to a path length cutoff $\Lambda$, which we choose such that $1 - \sum_{\ell = 1}^\Lambda \rho(\ell) < 10^{-6}$.  Note that the calculations for the state-dependent quantities $\psi(\s)$ and $\tau(\s)$ are simplified in this model (compared to more general cases~\cite{Manhart2013, Manhart2014}) since the strong-selection dynamics prevents the population from traversing loops in genotype space.  The time complexity of the algorithm scales as $\mathcal{O}(\gamma N\Lambda)$~\cite{Manhart2013}, where $\gamma$ is the average connectivity and $N$ is the total size of the state space.  For genotypic sequences of length $L$ and an alphabet of size $k$, $\gamma \sim L(k-1)$ and $N \sim k^L$.


     \textbf{Validity of the additive energy model.}  Double mutant experiments indicate that the additive energy model is a good approximation for residues that are not in direct physical contact~\cite{Wells1990, Istomin2008}.  For spatially-close residues, the mutational effects are largely ``sub-additive'' (diminishing-returns magnitude epistasis): two (de)stabilizing mutations combined will still usually be (de)stabilizing, but less so than the sum of their individual effects~\cite{Wells1990, Istomin2008}.  For example, Istomin et al.~\cite{Istomin2008} find that while residues separated by more than 6 \AA~are nearly additive (correlation $R^2=0.97$ with a slope of $0.88$ between the sum of $\Delta \Delta G$'s for two single mutants and $\Delta \Delta G$ for the double mutant), spatially-close residues are substantially sub-additive ($R^2=0.84$, slope of $0.54$).  Nonetheless, in regions with straight contours which represent most of our fitness landscapes, sub-additive energies cannot produce sign epistasis; substantial deviations from energy sub-additivity are required to create additional local maxima or place significant constraints on adaptive paths. Thus it appears that deviations from energy additivity will not lead to qualitative changes in our model's predictions.


%
%
%
%
%
%
%
%


\newpage


\renewcommand{\thefigure}{S\arabic{figure}}
\setcounter{figure}{0}

\section*{Supplementary Figures}

\begin{figure}[ht!]
\centering\includegraphics[scale=0.90]{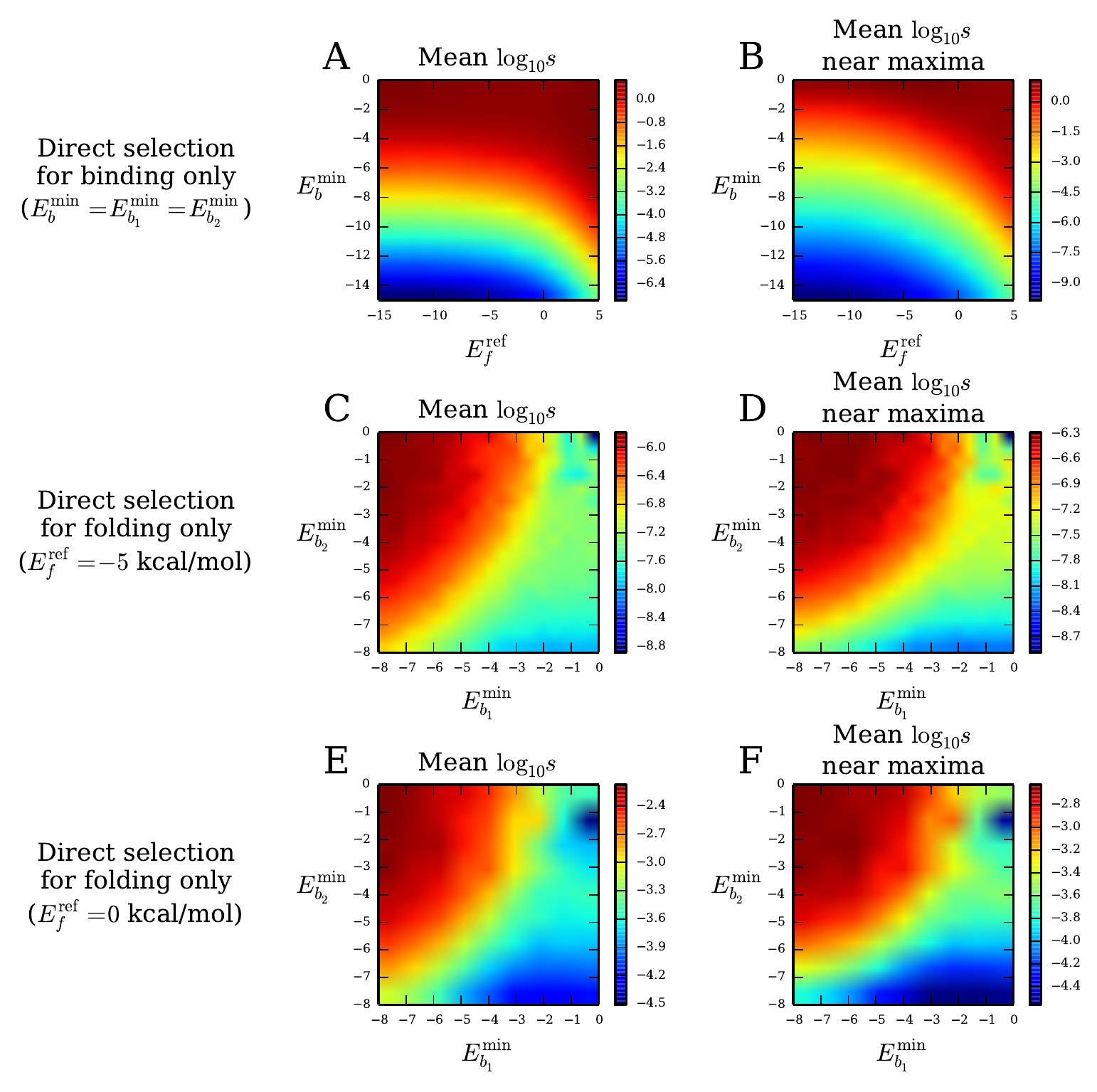}
\caption{
\textbf{Average selection strength.}  
(A)~Average $\log_{10} s$ ($s$ is the selection coefficient) of all accessible beneficial substitutions as a function of $\EOf$ and $\EOb = \EObone = \EObtwo$ in the case of direct selection for binding only ($\fuf = \fub = 0$).  Due to the $E_b$ symmetry of this case (Fig.~1B), we can neglect differences in $\EObone$ and $\EObtwo$ without loss of generality. 
(B)~Same as (A) but limited to accessible substitutions that end at local fitness maxima.  
(C)~Average $\log_{10} s$ of all accessible beneficial substitutions as a function of
$\EObone$ and $\EObtwo$ in the case of selection for folding only ($\fuf = 0$, $\fub = 1$, $\EOf = -5$ kcal/mol).
(D)~Same as (C) but limited to accessible substitutions that end at local fitness maxima.
(E, F)~Same as (C, D) but for $\EOf = 0$ kcal/mol.
Simultaneous selection for both binding and folding yields
qualitatively similar results.
All data points are averages over $10^4$ landscape realizations.
}
\label{fig:selection_strength}
\end{figure}

\clearpage

\begin{figure}[ht!]
\centering\includegraphics[scale=1.0]{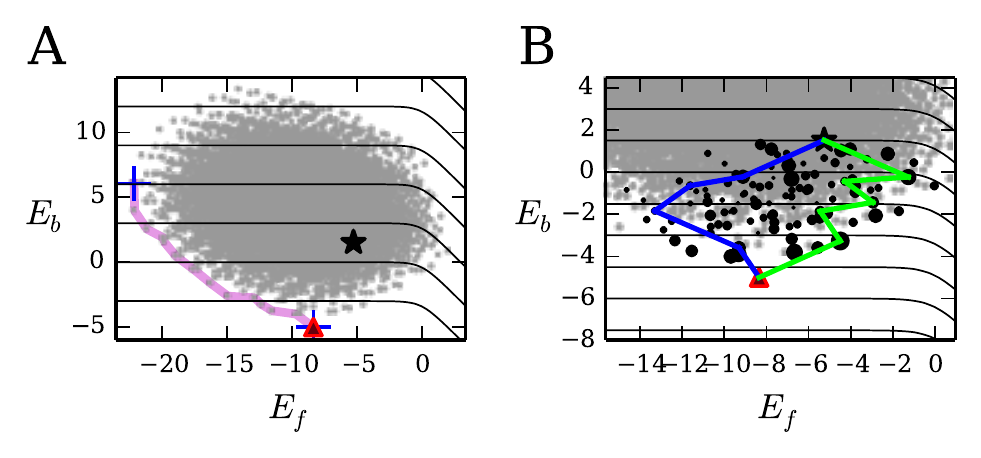}
\centering\includegraphics[scale=1.0]{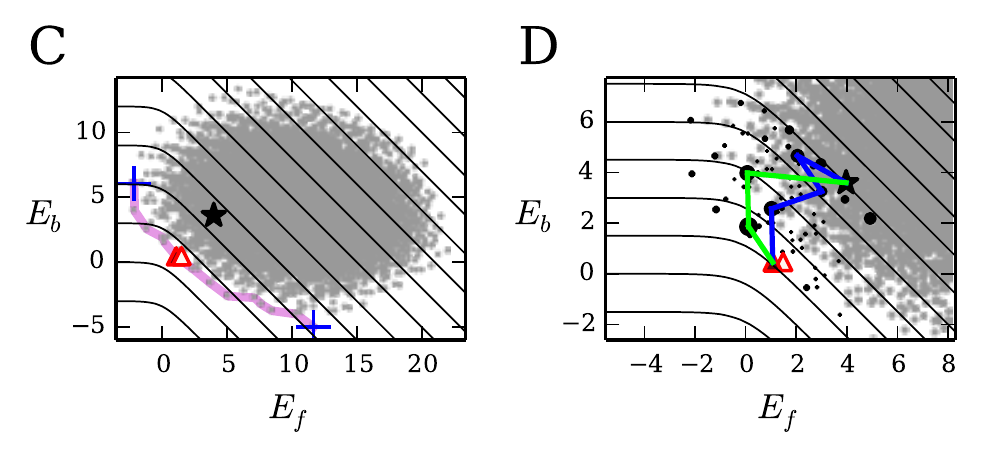}
\caption{
\textbf{Example landscapes for stable and intrinsically unstable proteins with direct selection for binding only.}  Symbols and randomly generated energy matrices ($\epsilon_f$, $\epsilon_{b_1}$, and $\epsilon_{b_2}$) are the same as in Fig.~2A,B.  
(A, B)~Stable protein ($\EOf = -15$ kcal/mol). 
(C, D)~Intrinsically unstable protein ($\EOf = 5$ kcal/mol). 
As in Fig.~2A,B, $\fub = \fuf = 0$ and $\EObone = \EObtwo = -5$ kcal/mol. 
}
\label{fig:extra_case1_examples}
\end{figure}

\newpage

\begin{figure}[ht!]
\centering\includegraphics[scale=1.0]{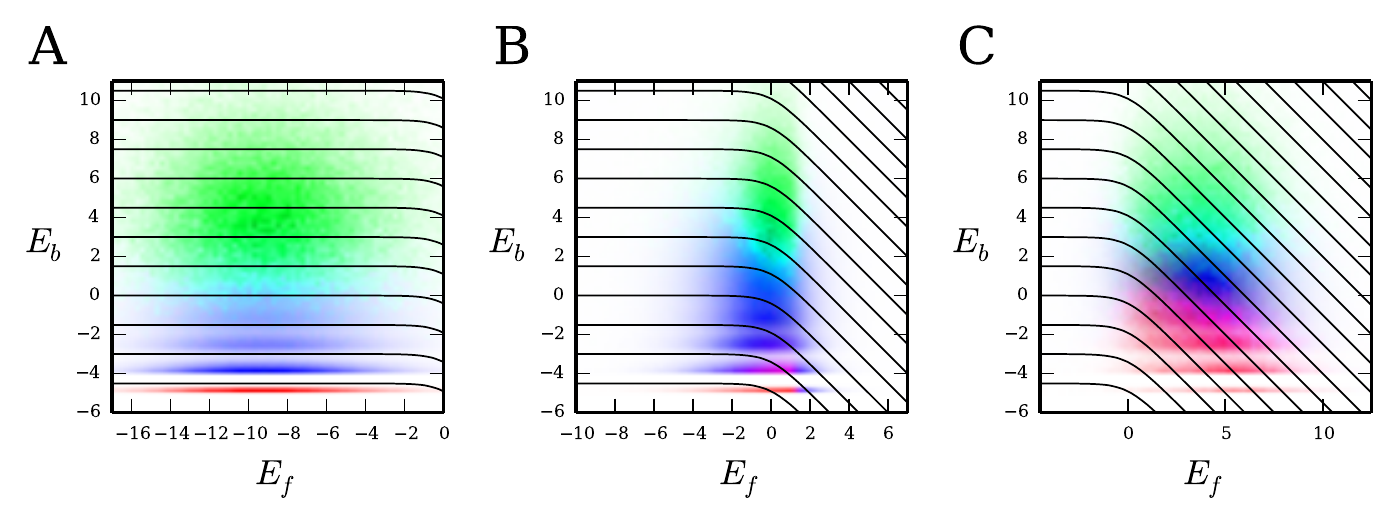}
\caption{
\textbf{Average landscapes for direct selection for binding only.}  
As in Fig.~4C, the distribution of initial states is shown in green, intermediate states in blue (weighted by their path densities), and final states in red (weighted by their commitment probabilities).  
(A)~Stable proteins ($\EOf = -15$ kcal/mol).
(B)~Marginally stable proteins ($\EOf = -3$ kcal/mol). 
(C)~Intrinsically unstable proteins ($\EOf = 5$ kcal/mol).
All landscapes are averaged over $10^5$ realizations.  As in Fig.~2A,B, $\fub = \fuf = 0$ and $\EObone = \EObtwo = -5$ kcal/mol. 
}
\label{fig:case1_landscape_averages}
\end{figure}

\begin{figure}[ht!]
\centering
\includegraphics[scale=1.0]{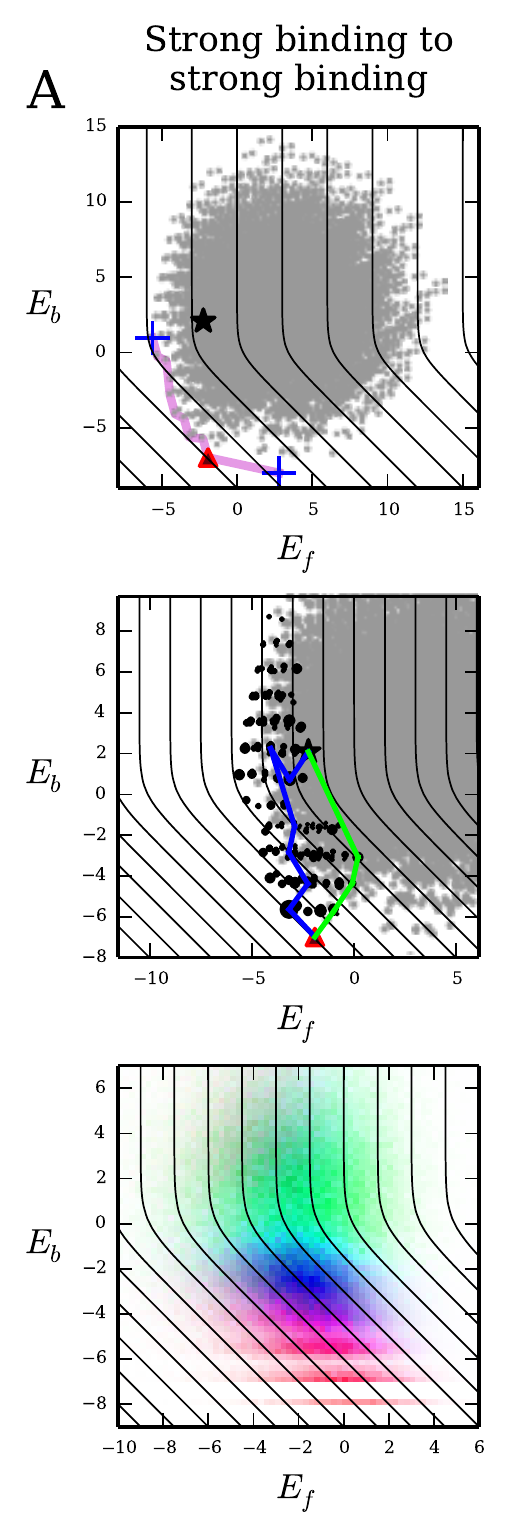}
\includegraphics[scale=1.0]{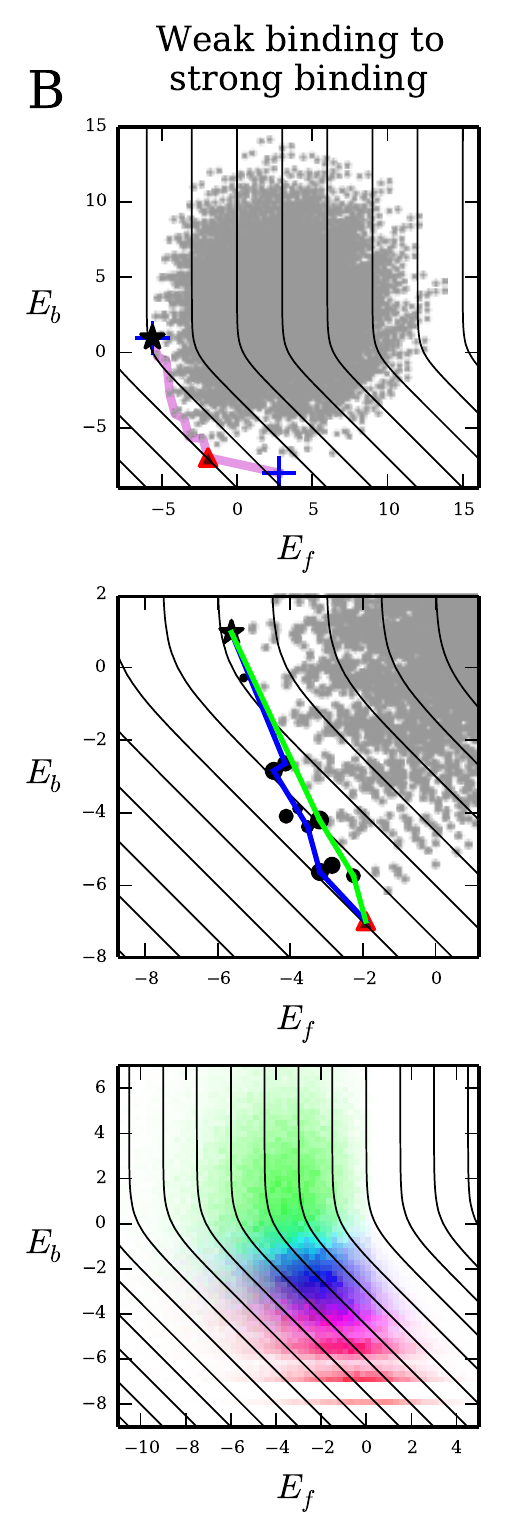}
\includegraphics[scale=1.0]{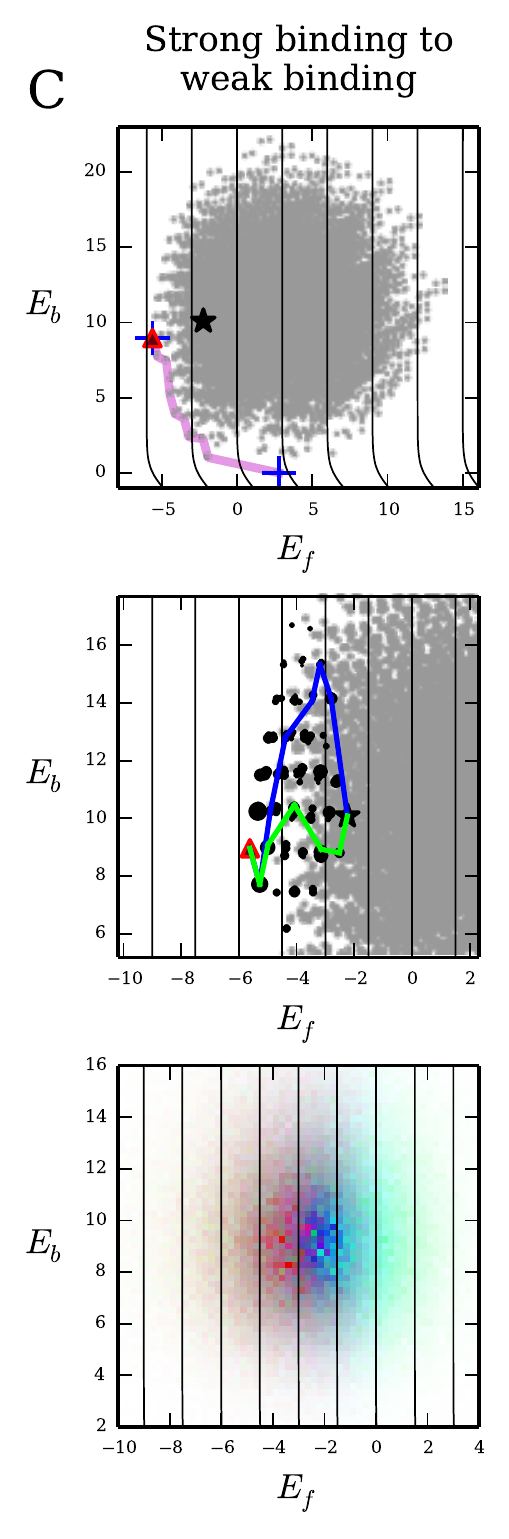}
\caption{
\textbf{Example and average landscapes for direct selection for folding only.}  
Symbols in top and middle panels are the same as in Fig.~2A,B, and the color scheme in the bottom panels is the same as in Fig.~4C and Fig.~\ref{fig:case1_landscape_averages}.  
(A)~Strong binding to both old and new targets ($\EObone = \EObtwo = -8$ kcal/mol). 
(B)~Weak binding to old target and strong binding to new target ($\EObone = 0$ kcal/mol, $\EObtwo = -8$ kcal/mol). 
(C)~Strong binding to old target and weak binding to new target ($\EObone = -8$ kcal/mol, $\EObtwo = 0$ kcal/mol). 
We use $\fub = 1$, $\fuf = 0$, and $\EOf = 0$ kcal/mol in all cases.  In the bottom panels, the landscapes are averaged over $10^5$ realizations.
}
\label{fig:case2_landscapes}
\end{figure}

\begin{figure}[ht!]
\centering\includegraphics[scale=1.0]{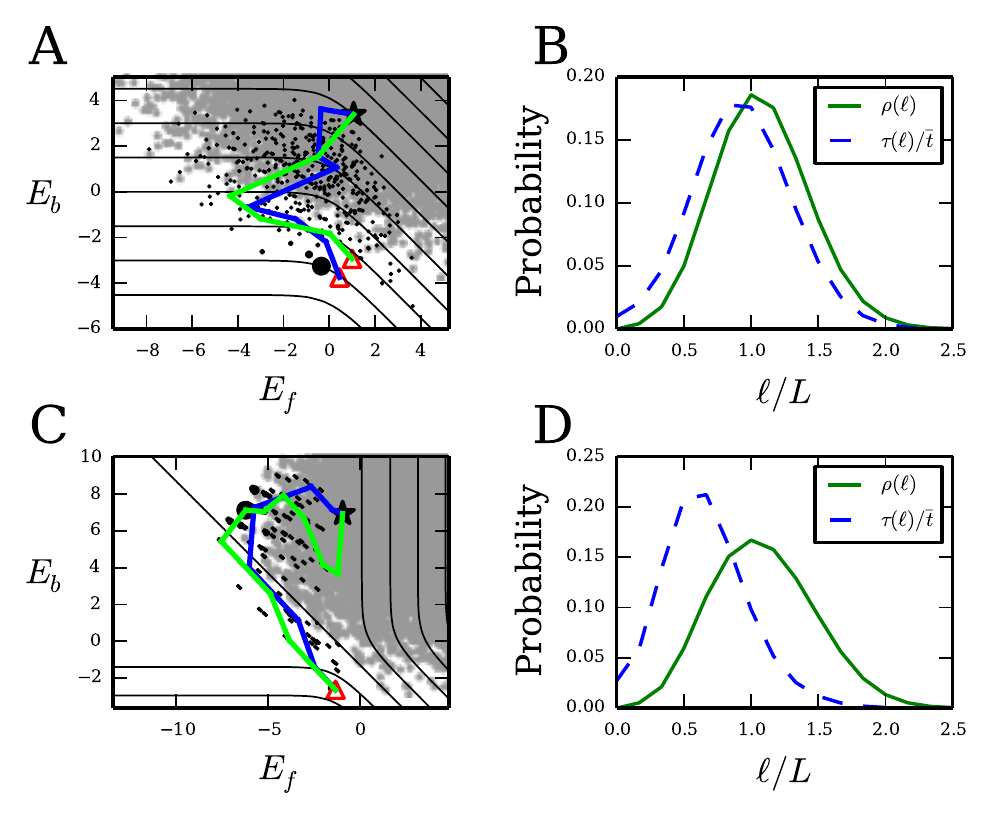}
\caption{
\textbf{Distribution of adaptation times over intermediate states.}  
(A)~The same landscape realization as in Fig.~2A,B (selection for binding only on a marginally stable protein), but with each intermediate state $\s$ sized proportional to $\tau(\s)$, the average time spent in that state. 
(B)~The probability $\rho(\ell)$ (solid, green) of taking an adaptive path of exactly $\ell$ substitutions and the average time $\tau(\ell)$ (dashed, blue) spent by paths at the $\ell$th substitution, averaged over $10^5$ realizations with $\fub = \fuf = 0$, $\EOf = -3$ kcal/mol, and $\EObone = \EObtwo = -5$ kcal/mol. 
(C, D)~Same as (A, B), but with the landscape realization used in Fig.~4A,B (selection for both binding and folding, $\fub = 0.9$, $\fuf = 0$, $\EOf = \EObone = \EObtwo = -4$ kcal/mol).
}
\label{fig:tempo}
\end{figure}

\end{document}